\documentclass[aps,twocolumn,showpacs]{revtex4}
\usepackage{amsmath}
\usepackage{epsfig}

\begin{document}

\title{Are $N\bar{\Omega}$ bound states?}

\author{Hongxia Huang$^1$\footnote{Email: hxhuang@njnu.edu.cn},
Xinmei Zhu$^2$\footnote{Email: zxm\_yz@126.com},
Jialun Ping$^1$\footnote{Corresponding author: jlping@njnu.edu.cn},
Fan Wang$^3$\footnote{Email: fgwang@foxmail.com},
T. Goldman$^4$\footnote{Email: tgoldman@lanl.gov}}

\affiliation{$^1$Department of Physics, Nanjing Normal University,
Nanjing 210097, P.R. China}

\affiliation{$^2$Department of Physics, Yangzhou University,
Yangzhou 225009, P.R. China}

\affiliation{$^3$Department of Physics, Nanjing University,
Nanjing 210093, P.R. China}

\affiliation{$^4$Department of Physics and Astronomy, University of New Mexico, Albuquerque, NM 87501}

\begin{abstract}
Inspired by the progress of the experimental search of the $N\Omega$ dibaryon by the STAR collaboration, we study $N\bar{\Omega}$
systems in the framework of quark delocalization color screening model. Our results show that the attraction between $N$ and $\bar{\Omega}$
is a little bit larger than that between $N$ and $\Omega$, which indicates that it is more possible for the $N\bar{\Omega}$ than the
$N\Omega$ system to form bound states. The dynamic calculations state that both the $J^{P}=1^{+}$ and $2^{+}$ $N\bar{\Omega}$ systems
are bound states. The binding energy of these two states are deeper than that of $N\Omega$ systems with $J^{P}=2^{+}$, and the $N\Omega$
system with $J^{P}=1^{+}$ is unbound. The calculation of the low-energy scattering phase shifts, scattering length and the effective range
also supports the existence of the $N\bar{\Omega}$ bound states with $J^{P}=1^{+}$ and $2^{+}$. So the $N\bar{\Omega}$ states are better
hexaquark states and stronger signals are expected in experiments.
\end{abstract}


\maketitle

\setcounter{totalnumber}{5}

\section{\label{sec:introduction}Introduction}
The studies of baryon-antibaryon bound states can be backdated to the proposal of Fermi and Yang~\cite{fermi} to make the pion with a
nucleon-antinucleon ($N\bar{N}$) pair. In the traditional one-boson-exchange theory of
nucleon-nucleon ($NN$) interaction, it is shown that $N\bar{N}$ system is more attractive than $NN$ system due to the strong
$\omega$-exchange~\cite{PR13}. Therefore, possible bound states or resonances of a nucleon-antinucleon system have
been proposed for many years. An extensive and excellent review of the possible bound states of $N\bar{N}$ was given in
Ref.~\cite{richard}.

Recently, the experimental progresses on dibaryons renewed our concerns for dibaryon systems. The WASA@COSY collaboration reported
the discovery of dibaryon $d^*$~\cite{dstar}, which was predicted by Xuong and Dyson in 1964~\cite{d03} and was named as an 
``inevitable" dibaryon in 1989~\cite{inevitable}. The discovery was confirmed recently by A2 collaboration at MAMI~\cite{MAMI}.
In 2019, the STAR collaboration measured the $p\Omega$ correlation function in $Au+Au$ collisions at the Relativistic Heavy-Ion 
Collider (RHIC)~\cite{RHIC}, and reported that the scattering length is positive for the $p\Omega$ interaction and favored 
the $p\Omega$ bound state hypothesis.
The $S=-3$, $I=1/2$, $J=2$ $N\Omega$ state was predicted by Goldman {\em et al.} as a narrow resonance in a relativistic
quark model~\cite{PRL59}. Oka also proposed that there should be a quasi-bound state with $IJ^{P}=\frac{1}{2}2^{+}$ by using
a constituent quark model~\cite{Oka}. Recent study of the lattice QCD by HAL QCD Collaboration reported that the $N\Omega$ was
indeed a bound state at pion mass of $875$ MeV~\cite{HAL1} and later with nearly physical quark masses ($m_{\pi}\simeq 146$ MeV
and $m_{K}\simeq 525$ MeV)~\cite{HAL2}. Then, by employing the interactions obtained from the ($2+1$)-flavor lattice QCD simulations,
K. Morita {\em et al.} studied the two-pair momentum correlation functions of the $N\Omega$ state in relativistic heavy-ion collisions
to investigate the existence of $N\Omega$ bound state~\cite{Morita1,Morita2}. Our group has showed that the $N\Omega$ with
$I=1/2$, $J=2$ was a narrow resonance state~\cite{PRC51,PangHR}. We also investigated this state in the $D-$wave $\Lambda\Xi$ scattering
phase shifts and found that the $N\Omega$ was really a narrow resonance in the $D-$wave $\Lambda\Xi$ scattering process~\cite{ChenM}.
The low-energy scattering phase shifts, scattering length, and effective range of the $N\Omega$ system were also calculated to confirm
the existence of $N\Omega$ resonance~\cite{Huang1}. Besides, this state has also been observed to be bound in the chiral quark model~\cite{LiQB}.

By analogy to the $NN$ and $N\bar{N}$ systems, one may speculate that there should be a bit more attraction in the $N\bar{\Omega}$ channel
than the one in the $N\Omega$ channel. If the $N\Omega$ resonance can be confirmed in further study, we hope an even stronger signal of
the $N\bar{\Omega}$ state can be found in experiments. Besides, the $N\bar{N}$ state would annihilate very quickly in the ground state
because of the quark contents of this system, so it is still difficult to give a convincing theoretical confirmation of the $N\bar{N}$
bound state or resonance. While for the $N\bar{\Omega}$ state, it cannot annihilate to the vacuum, since $N$ is composed of three $u$($d$)
quarks and $\bar{\Omega}$ of three $\bar{s}$ quarks. Therefore, the $N\bar{\Omega}$ state is a more stable and special state, which can provide
an optimal place to study the baryon-antibaryon interactions. The copious production of anti-baryons in the high energy collider provides good
opportunities to study this type of spectrum. Clearly, the theoretical study of the $N\bar{\Omega}$ system is interesting and necessary,
which can provide more information for the experimental search for the baryon-antibaryon bound states.

It is well known that the forces between nucleons (hadronic clusters of quarks) are qualitative similar to the forces between atoms
(clusters of charged particles). One of the constituent quark models, quark delocalization color screening model (QDCSM)~\cite{QDCSM1}
is this molecular model of nuclear forces. Two new ingredients are introduced in this model: the one is the quark delocalization, which
is used to consider the orbital excitation by allowing quarks delocalize from one cluster to another cluster; another one is the color
screening factor, which is used to modify the confinement interaction between quarks in different cluster orbits. In the study of $NN$
and $YN$ interactions and the properties of deuteron, the mechanism of quark delocalization and color screening is really responsible
for the intermediate range attraction~\cite{QDCSM2}. This model has also been employed to study some dibaryon candidates, such as
$d^{*}$~\cite{Ping1}, $N\Omega$~\cite{Huang1} and so on. It has also been extended to the baryon-antibaryon systems, like $p\bar{p}$
and $p\bar{\Lambda}$ systems~\cite{Huang2,Huang3}. Extending to the $N\bar{\Omega}$ system is natural. So we continue to investigate
the $N\bar{\Omega}$ system within the QDCSM in this work. By comparing with the interaction of the $N\Omega$ system, we will explore
whether the attraction in the $N\bar{\Omega}$ system is larger than the one in the $N\Omega$ system, and whether it is more possible
to form any bound state in the $N\bar{\Omega}$ system.

The structure of this paper is as follows. A brief introduction of the QDCSM is given in section II. Section III devotes to the numerical
results and discussions. The summary is shown in the last section.

\section{The quark delocalization color screening
model (QDCSM)}
The detail of QDCSM used in the present work can be found in the
references~\cite{QDCSM1,QDCSM2}. Here, we
just present the salient features of the model. The model
Hamiltonian is:
\begin{widetext}
\begin{eqnarray}
H &=& \sum_{i=1}^6 \left(m_i+\frac{p_i^2}{2m_i}\right) -T_c + \sum_{i<j=1}^3 V_{qq}(\mathbf{r}_{ij})
+ \sum_{i<j=4}^6 V_{\bar{q}\bar{q}}(\mathbf{r}_{ij})+ \sum_{i=1,3,j=4,6} V_{q\bar{q}}(\mathbf{r}_{ij}),
\label{E1}
\end{eqnarray}
where $T_c$ is the kinetic energy of the center of mass, and $V_{qq}(r_{ij})$ stands for the interaction between two quarks,
\begin{eqnarray}
V_{qq}(\mathbf{r}_{ij})&=& V_{qq}^{C}(\mathbf{r}_{ij})+V_{qq}^{G}(\mathbf{r}_{ij})+V_{qq}^{\chi}(\mathbf{r}_{ij}),   \\
\label{E3}
V_{qq}^{C}(\mathbf{r}_{ij})&=& -a_c \boldsymbol{\lambda}_i \cdot \boldsymbol{\lambda}_j [f(\mathbf{r}_{ij})+V_0], \\
 f(\mathbf{r}_{ij}) & = &  \left\{ \begin{array}{ll}
 r_{ij}^2 &
 \qquad \mbox{if }i,j\mbox{ occur in the same baryon orbit} \\
  \frac{1 - e^{-\mu_{ij} r_{ij}^2} }{\mu_{ij}} & \qquad
 \mbox{if }i,j\mbox{ occur in different baryon orbits} \\
 \end{array} \right.
 \\
\label{E5}
V_{qq}^{G}(\mathbf{r}_{ij})&=& \frac{1}{4}\alpha_{s} \boldsymbol{\lambda}_i \cdot
\boldsymbol{\lambda}_j
\left[\frac{1}{r_{ij}}-\frac{\pi}{2}\left(\frac{1}{m_{i}^{2}}+\frac{1}{m_{j}^{2}}+\frac{4\boldsymbol{\sigma}_i
\cdot \boldsymbol{\sigma}_j}{3m_{i}m_{j}}
 \right)
\delta(\mathbf{r}_{ij})-\frac{3}{4m_{i}m_{j}r^3_{ij}}S_{ij}\right],
 \\
\label{E6}
V_{qq}^{\chi}(\mathbf{r}_{ij})&=& \frac{1}{3}\alpha_{ch}
\frac{\Lambda^2}{\Lambda^2-m_{\chi}^2}m_\chi \left\{ \left[
Y(m_\chi r_{ij})- \frac{\Lambda^3}{m_{\chi}^3}Y(\Lambda r_{ij})
\right]
\boldsymbol{\sigma}_i \cdot \boldsymbol{\sigma}_j \right. \nonumber \\
&& \left. +\left[ H(m_\chi r_{ij})-\frac{\Lambda^3}{m_\chi^3}
H(\Lambda r_{ij})\right] S_{ij} \right\} \boldsymbol{\lambda}^f_i \cdot
\boldsymbol{\lambda}^f_j, ~~~\chi=\pi,K,\eta \\
S_{ij} & = &  \frac{(\boldsymbol{\sigma}_i \cdot {\mathbf r}_{ij})
(\boldsymbol{\sigma}_j \cdot \mathbf{r}_{ij})}{r_{ij}^2}-\frac{1}{3}~\boldsymbol{\sigma}_i \cdot \boldsymbol{\sigma}_j.
\end{eqnarray}
\end{widetext}
where $V_{qq}^{C}(r_{ij})$, $V_{qq}^{G}(r_{ij})$, and $V_{qq}^{\chi}(r_{ij})$ are the the confinement interactions,
one-gluon-exchange, and one-boson-exchange; $S_{ij}$ is quark tensor operator; $Y(x)$ and $H(x)$ are standard
Yukawa functions; $\alpha_{ch}$ is the chiral coupling constant, determined as usual from the $\pi$-nucleon coupling constant;
$\alpha_{s}$ is the quark-gluon coupling constant. In order to cover the wide energy range from light to strange quarks,
an effective scale-dependent quark-gluon coupling $\alpha_{s}(u)$ was introduced \cite{Vijande}:
\begin{eqnarray}
\alpha_{s}(u) & = &
\frac{\alpha_{0}}{\ln(\frac{u^2+u_{0}^2}{\Lambda_{0}^2})}.
\end{eqnarray}
The other symbols in the above expressions have their usual meanings. $V_{\bar{q}\bar{q}}(r_{ij})$ and $V_{q\bar{q}}(r_{ij})$
in Eq.~(\ref{E1}) represent the antiquark-antiquark ($\bar{q}\bar{q}$) and quark-antiquark ($q\bar{q}$) interactions.
\begin{eqnarray}
V_{\bar{q}\bar{q}}(r_{ij})&=& V_{\bar{q}\bar{q}}^{C}(r_{ij})+V_{\bar{q}\bar{q}}^{G}(r_{ij})+V_{\bar{q}\bar{q}}^{\chi}(r_{ij}),
\end{eqnarray}
and
\begin{eqnarray}
V_{q\bar{q}}(r_{ij})&=& V_{q\bar{q}}^{C}(r_{ij})+V_{q\bar{q}}^{G}(r_{ij})+V_{q\bar{q}}^{\chi}(r_{ij}),,
\end{eqnarray}
For the antiquark, replacing the $\boldsymbol{\lambda}_i$ in Eqs.~(\ref{E3}) and (\ref{E5}) by $-\boldsymbol{\lambda}^{*}_i$,
$\boldsymbol{\lambda}^f_i$ in Eq.~(\ref{E6}) by $\boldsymbol{\lambda}^{f*}_i$, the form of $V_{\bar{q}\bar{q}}$ and
$V_{q\bar{q}}$ can be obtained. It is noted that there is no annihilation between quark and antiquark. The reason is that
the $N\bar{\Omega}$ state cannot annihilate to the vacuum due to the different quark flavor contents of $N$ and $\bar{\Omega}$.
All parameters, which are fixed by fitting to the masses of the ground baryons, are taken from our previous work of $N\Omega$
system~\cite{Huang1}.

\section{The results and discussions}
The $S$-wave $N\bar{\Omega}$ systems with isospin $I=\frac{1}{2}$, spin parity $J^{P}=1^{+}$ and $J=2^{+}$ are investigated
in QDCSM. As the first step, we calculate the effective potentials of the $N\bar{\Omega}$ system, since an attractive potential
is necessary for forming bound state or resonance. The effective potential between two hadrons is defined as: $V(s)=E(s)-E(\infty)$,
where $E(s)$ is the energy of the system in the adiabatic approximation. $s$ is the generating coordinate, the separation between
two reference centers which two clusters are fixed to. The effective potentials of $J^{P}=1^{+}$ and $J=2^{+}$ are shown in Fig. 1
and Fig. 2, respectively. In order to compare with the interaction between $N$ and $\Omega$, the potentials of the $N\Omega$ channel
are also shown in two figures. Besides, the contributions of all terms of interactions to the effective potentials are also shown
in both two figures.

\begin{figure}
\epsfxsize=3.6in \epsfbox{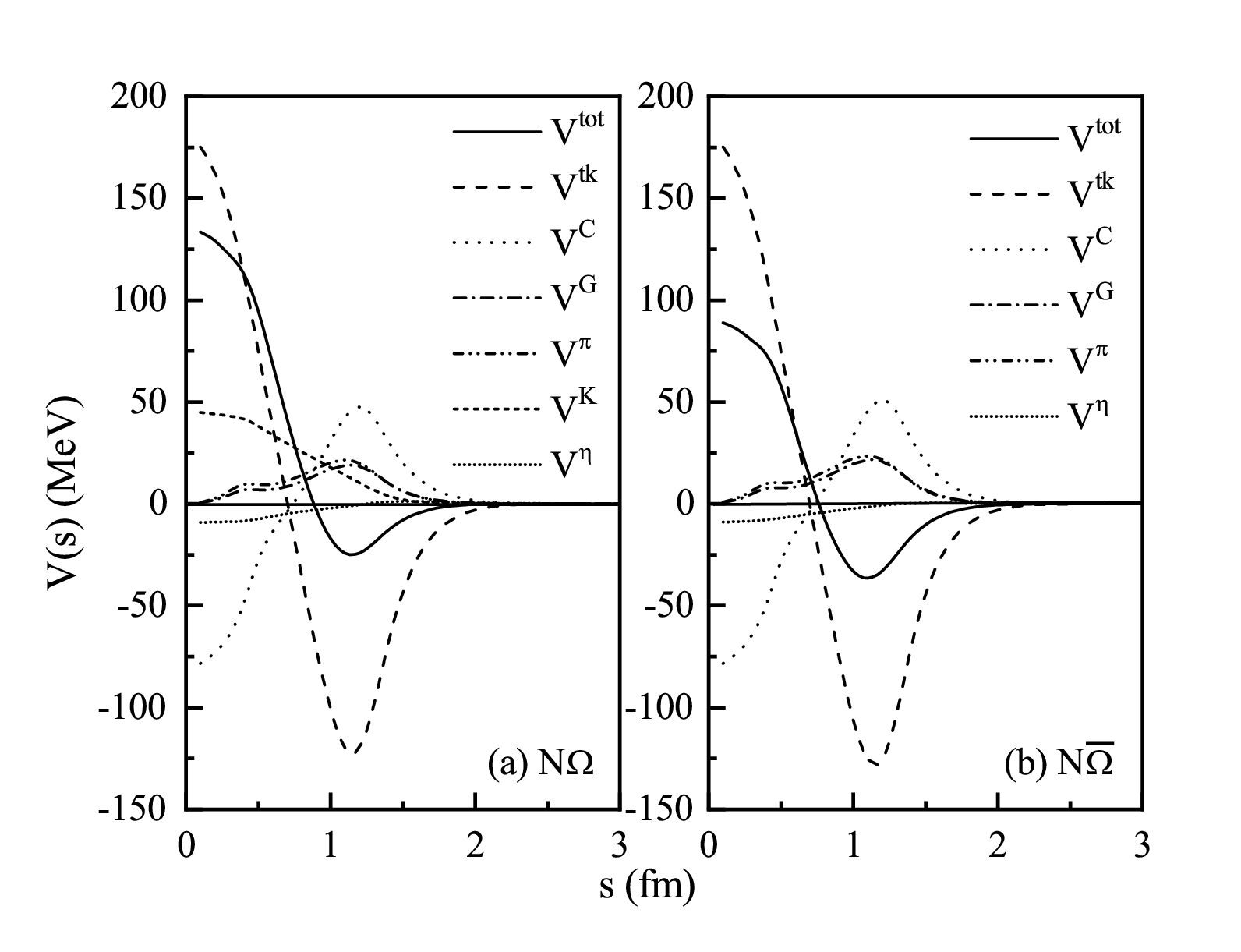} \vspace{-0.1in}

\caption{The effective potentials of $N\Omega$ and $N\bar{\Omega}$ systems with $J^{P}=1^{+}$.}
\end{figure}

\begin{figure}
\epsfxsize=3.6in \epsfbox{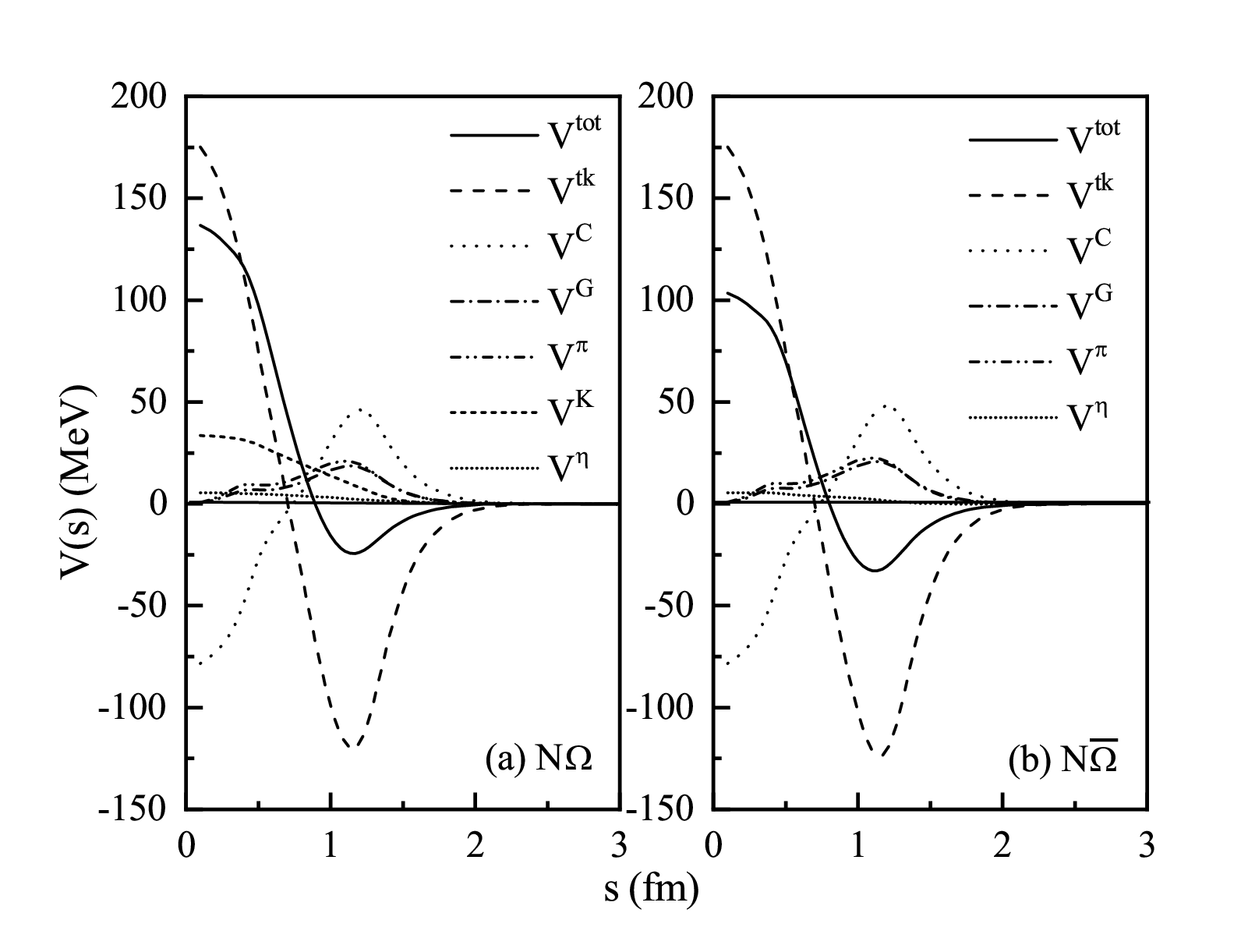} \vspace{-0.1in}

\caption{The effective potentials of $N\Omega$ and $N\bar{\Omega}$ systems with $J^{P}=2^{+}$.}
\end{figure}

From the two figures, we can see that the potentials are attractive for both the $J^{P}=1^{+}$ and $J=2^{+}$ $N\bar{\Omega}$ systems.
The attraction of the $N\bar{\Omega}$ channel with $J^{P}=1^{+}$ is a little bit larger than the one with $J^{P}=2^{+}$. By comparing
with the potential of the $N\Omega$ channel, the interaction between $N$ and $\bar{\Omega}$ is more attractive than that between $N$
and $\Omega$, which indicates that it is more possible for the $N\bar{\Omega}$ system than the $N\Omega$ system to form a bound state.
This rule is similar with the conclusion of $NN$ and $N\bar{N}$ systems in Ref.~\cite{PR13}, which points out that $N\bar{N}$ is more
attractive than $NN$ system.

The contributions to the effective potential of each term in the system hamiltonian, the kinetic energy ($V^{tk}$), the confinement ($V^{C}$),
the one-gluon-exchange ($V^{G}$), and the one-boson-exchange ($V^{\pi}$, $V^{K}$, and $V^{\eta}$), are shown in both two figures.
From the figures, one can see that quark delocalization and color screening work together to provide appropriate short-range repulsion and
intermediate-range attraction. For the $J^{P}=1^{+}$ system, it is obvious that the attraction of both the $N\Omega$ and $N\bar{\Omega}$ systems
mainly comes from the kinetic energy term (due to the quark delocalization), and other terms provide repulsive potentials, which reduce the
total attraction of the potentials.
The only difference is that the attraction of the kinetic energy term of the $N\bar{\Omega}$ channel is a little larger than that of the
$N\Omega$ channel. Besides, one notes that the $K$-meson exchange interactions ($V^{K}$) do not contribute at all to the effective potential
between $N$ and $\bar{\Omega}$ due to the quark contents of these two hadrons. For the $J^{P}=2^{+}$ system, the results are similar.
Only the attraction of the kinetic energy term of the $J^{P}=2^{+}$ system is a little smaller than that of the $J^{P}=1^{+}$ system.

In order to see whether or not there is any bound state, a dynamic calculation is needed. Here we use the resonating group method (RGM)
to solve a bound-state problem, which was described in more detail in Ref.~\cite{RGM}. Expanding the relative motion wave function
between two hadrons in the RGM equation by gaussians, the integro-differential equation of RGM can be reduced to an algebraic equation,
the generalized eigen-equation. Then the energy of the system can be obtained by solving the eigen-equation. In order to keep the matrix
dimension manageably small, the hadron-hadron separation is taken to be less than 6 fm in our calculation. The binding energies of the
$N\bar{\Omega}$ system with both $J^{P}=1^{+}$ and $J^{P}=2^{+}$, compared with the result of the $N\Omega$ system are listed in
Table~\ref{bound}, where $B_{sc}$ stands for the binding energy of the single channel $N\Omega$/$N\bar{\Omega}$, and $B_{cc}$ means the
binding energy of the channel-coupling. There is only one channel for the $N\bar{\Omega}$ state, because we limit our study of the system
to color-singlet sub-clusters, three $u/d$ quarks and three $\bar{s}$ quarks here. While for the $N\Omega$ state with three $u/d$ quarks
and three $s$ quarks, there are seven color-singlet channels for the $J^{P}=1^{+}$ system, including $\Xi\Sigma$, $\Xi\Lambda$,
$\Xi^{*}\Sigma$, $\Xi\Sigma^{*}$, $\Xi^{*}\Lambda$, $N\Omega$, and $\Xi^{*}\Sigma^{*}$; and five channels for the $J^{P}=2^{+}$ system,
including $\Xi^{*}\Sigma$, $\Xi\Sigma^{*}$, $\Xi^{*}\Lambda$, $N\Omega$, and $\Xi^{*}\Sigma^{*}$.

\begin{table}
\caption{The binding energy (in MeV) of both the $N\Omega$ and $N\bar{\Omega}$ systems.}
\begin{tabular}{cccc}
\hline \hline
   &\multicolumn{2}{c}{\rm $N\Omega$}&\multicolumn{1}{c}{\rm $N\bar{\Omega}$}\\ \hline
 ~~~~$J^{P}$~~~~ & ~~~~$B_{sc}$~~~~  & ~~~~$B_{cc}$~~~~ & ~~~~$B_{sc}$~~~~  \\ \hline
    $~ 1^{+}$  & $ub$ & $ub$ & $-10.4$    \\
    $~ 2^{+}$  & $ub$ & $-6.4$ & $-8.7$  \\
\hline \hline
\end{tabular}
\label{bound}
\end{table}

From Table~\ref{bound} we can see that both the $J^{P}=1^{+}$ and $J^{P}=2^{+}$ $N\bar{\Omega}$ are bound states with the binding energy of
$-10.4$ MeV and $-8.7$ MeV, respectively. By contrast, the single channel calculation shows that neither the $J^{P}=1^{+}$ nor $J^{P}=2^{+}$ $N\Omega$ is bound (labelled as '$ub$'
in Table~\ref{bound}). Only multi-channel coupling can make the $J^{P}=2^{+}$ $N\Omega$ to be bound with binding energy of $-6.4$ MeV.
All these results show that it is more possible for the $N\bar{\Omega}$ system rather than the $N\Omega$ system to form a bound state.
Therefore, if there are some signals for the $N\Omega$ resonance in experiments, we look forward an even stronger signal of the
$N\bar{\Omega}$ state can be found in experiments.

Generally, the bound state with the smaller spin should have lower mass. It is obvious that this order is normal in the $N\bar{\Omega}$
system. However, for the $N\Omega$ system, we only obtain a $N\Omega$ bound state with $J^{P}=2^{+}$. The reason is that the $N\Omega$
is the lowest channel in the $J^{P}=2^{+}$ system, it can be pushed down by other higher channels' coupling. While for the $J^{P}=1^{+}$
system, the lowest channel is the $\Xi\Lambda$, and the interaction between $\Xi$ and $\Lambda$ is repulsive. It cannot be lowered down
enough to form any bound state. And the $N\Omega$ channel maybe pushed down or drive up by other channels coupling. So there is no
$N\Omega$ bound state with $J^{P}=1^{+}$.

In our previous work of $N\Omega$ system with $IJ^{P}=\frac{1}{2}~ 2^{+}$, we also calculate the low-energy scattering phase shifts,
scattering length and the effective range to provide more theoretical input for the experiment~\cite{Huang1}. It is natural to do the
same work for the $N\bar{\Omega}$ system, as well as the $N\Omega$ system with $IJ^{P}=\frac{1}{2}~ 1^{+}$. We calculate the low-energy
scattering phase shifts by using the well developed Kohn-Hulthen-Kato(KHK) variational method, the details of which can be found in
Ref.~\cite{RGM}. The low-energy scattering phase shifts of both $N\Omega$ and $N\bar{\Omega}$ systems are shown in Fig. 3.
The data of the $N\Omega$ system with $IJ^{P}=\frac{1}{2}~ 2^{+}$ are from our former work~\cite{Huang1}.

Clearly, for the $N\bar{\Omega}$ systems with both $J^{P}=1^{+}$ and $J^{P}=2^{+}$, as well the $N\Omega$ system with $J^{P}=2^{+}$,
the scattering phase shifts go to $180$ degrees at $E_{c.m.}\sim 0$ and rapidly decreases as $E_{c.m.}$ increases, which indicates
the existence of a bound state in these systems. Whereas the scattering phase shift of the $J^{P}=1^{+}$ $N\Omega$ system is close
to $0$ degrees at $E_{c.m.}\sim 0$, which implies the unbound status for this state. All these results are consistent with the
bound state calculation shown above.

\begin{figure}
\epsfxsize=3.6in \epsfbox{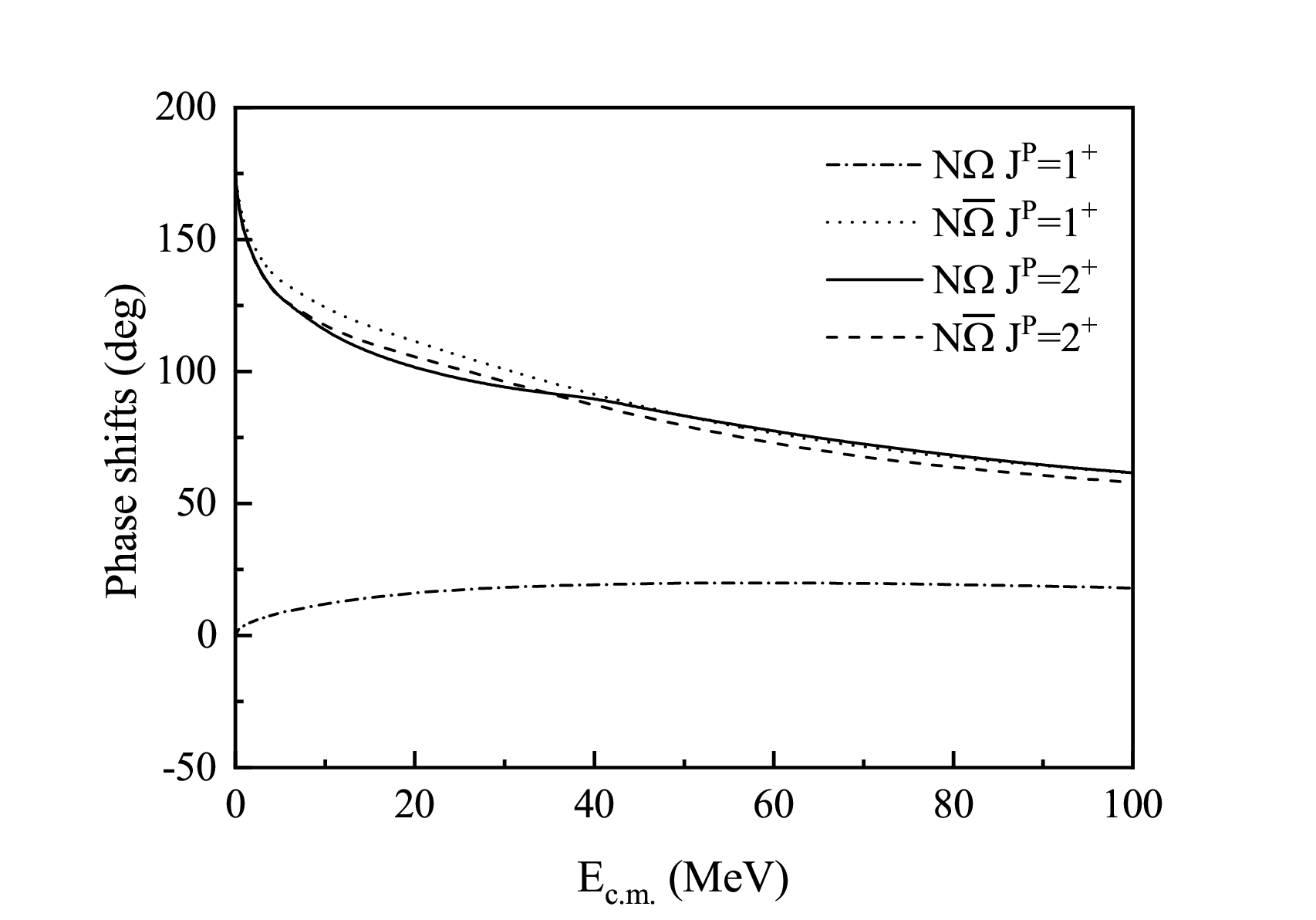} \vspace{-0.1in}

\caption{The phase shifts of $N\Omega$ and $N\bar{\Omega}$ systems.}
\end{figure}

Then, we can extract the scattering length $a_{0}$ and the effective range $r_{0}$ of the $N\Omega$ and $N\bar{\Omega}$ systems from
the low-energy phase shifts obtained above by using the expansion:
\begin{eqnarray}
k\, cot\delta & = & -\frac{1}{a_{0}}+\frac{1}{2}r_{0}k^{2}+{\cal
O}(k^{4})
\end{eqnarray}
where $k$ is the momentum of the relative motion with $k=\sqrt{2\mu E_{\mbox{c.m.}}}$, $\mu$ is the reduced mass of two baryons,
and $E_{\mbox{c.m.}}$ is the incident energy; $\delta$ is the low-energy scattering phase shifts. And the binding energy
$B^{\prime}$ can be calculated according to the following relation:
\begin{eqnarray}
B^{\prime} & = &\frac{\hbar^2\alpha^2}{2\mu}
\end{eqnarray}
where $\alpha$ is the wave number which can be obtained from the relation~\cite{Babenko}:
\begin{eqnarray}
r_{0} & = &\frac{2}{\alpha}(1-\frac{1}{\alpha a_{0}})
\end{eqnarray}
Please note that this is another method to calculate the binding energy, labelled as $B^{\prime}$.
The results are listed in Table~\ref{length}.
\begin{center}
\begin{table}[h]
\caption{The scattering length $a_{0}$, effective range $r_{0}$,
and binding energy $B^{\prime}$ of both the $N\Omega$ and $N\bar{\Omega}$ systems.}
\begin{tabular}{cccccccc}
\hline \hline
  ~~$J^{P}$~~ & ~~Channel~~ & ~~$a_{0}~(fm)$~~ & ~~$r_{0}~(fm)$~~ & ~~$B^{'}~(MeV)$~~    \\ \hline
  $ 1^{+}$ & $N\Omega$ & $-0.37807$  & $-2.3544$ & $ub$    \\
  & $N\bar{\Omega}$ & $2.4269$ & 0.47785 & $-6.9$   \\ \hline
 $ 2^{+}$ & $N\Omega$ & $2.8007$  & $0.5770$ & $-5.2$    \\
 & $N\bar{\Omega}$ & $2.7928$ & $0.81106$ & $-6.1$   \\
  \hline \hline
\end{tabular}
\label{length}
\end{table}
\end{center}

From Table~\ref{length}, we can see that the scattering length are positive for all systems except the $J^{P}=1^{+}$ $N\Omega$ system,
which confirms the existence of bound states for the $N\bar{\Omega}$ systems with $J^{P}=1^{+}$ and $2^{+}$, and the $N\Omega$ system
with $J^{P}=2^{+}$. Besides, the binding energies of these systems obtained by Eq. (12) is broadly consistent with that in
Table~\ref{bound}, which is obtained by the dynamic calculation.

\section{Summary}
In this work, we investigate the $S$-wave $N\bar{\Omega}$ systems with isospin $I=\frac{1}{2}$, spin parity $J^{P}=1^{+}$ and $2^{+}$
in the framework of QDCSM. To compare with the interaction between $N$ and $\Omega$, the $N\Omega$ systems with $J^{P}=1^{+}$ and
$2^{+}$ are also studied here. The results show that the $N\Omega$ systems with $J^{P}=1^{+}$ is unbound and $J^{P}=2^{+}$ is bound,
while both the $J^{P}=1^{+}$ and $2^{+}$ $N\bar{\Omega}$ systems are bound. The attraction between $N$ and $\bar{\Omega}$ is a little
bit larger than that between $N$ and $\Omega$, which indicates that the $N\bar{\Omega}$ is more possible to form bound state than the
$N\Omega$ state. If the $N\Omega$ resonance can be confirmed in experiments, we hope an even stronger signal of the $N\bar{\Omega}$
state can be found in experiments. The calculation of the low-energy scattering phase shifts, scattering length and the effective range
of the $N\bar{\Omega}$ systems also favors the existence of the $N\bar{\Omega}$ bound states with $J^{P}=1^{+}$ and $2^{+}$.
Besides, since $N$ is composed of three $u$($d$) quarks and $\bar{\Omega}$ of three $\bar{s}$ quarks, the $N\bar{\Omega}$ state cannot
annihilate to the vacuum. Therefore, the $N\bar{\Omega}$ state is a more stable and special state, which can provide more useful
information for the experimental search of the baryon-antibaryon bound states.

Our former work of non-strange dibaryons shows that the results in the QDCSM is consistent with that in the chiral quark model
(ChQM)~\cite{ChenLZ}, which means that the mechanism of intermediate-range attraction in two phenomenological quark models is
equivalent for the non-strange dibaryons. Extending to the strange dibaryons $N\Omega$, roughly identical results are obtained
in these two models~\cite{Huang1}. Here, by comparing with the work of Ref~\cite{ZhangD}, in which the $N\bar{\Omega}$ state with
spin $S=1$ and $S=2$ can be bound in both the chiral $SU(3)$ and the extended chiral $SU(3)$ quark models, the result of the
$N\bar{\Omega}$ state in QDCSM is also consistent with that in ChQM.
The mechanism of the intermediate-range attraction of the hadron-hadron interaction is one of the important issues in the quark model
study of hadron physics. The study of interactions between baryon and antibaryon here is also an effective place to test this mechanism.
More work is needed to explore the problem of the hadron-hadron interaction in future.

\acknowledgments{This work is supported partly by the National Science Foundation of China under
Contract Nos. 11175088, 11035006, 11205091.}

\end{document}